\documentclass[
    ,sort
    ,final            % use final for the camera ready runs
  ,numberedheadings % uncomment this option for numbered sections
  ]
  {aipproc}

\layoutstyle{6x9}

\begin{document}

\title{Entropic Priors and Bayesian Model Selection}


\author{Brendon J. Brewer\footnote{Email: \texttt{brewer@physics.ucsb.edu}}}{
  address={Department of Physics, University of California, Santa Barbara, CA, 93106-9530, USA}
}

\author{Matthew J. Francis}{
  address={SiSSA, via Beirut, 2-4 34151 Trieste TS, Italy}
}

\begin{abstract}
We demonstrate that the principle of maximum relative entropy (ME), used judiciously, can ease the specification of priors in model selection problems. The resulting effect is that models that make sharp predictions are disfavoured, weakening the usual Bayesian ``Occam's Razor''. This is illustrated with a simple example involving what Jaynes called a ``sure thing'' hypothesis. Jaynes' resolution of the situation involved introducing a large number of alternative ``sure thing'' hypotheses that were possible before we observed the data. However, in more complex situations, it may not be possible to explicitly enumerate large numbers of alternatives. The entropic priors formalism produces the desired result without modifying the hypothesis space or requiring explicit enumeration of alternatives; all that is required is a good model for the prior predictive distribution for the data. This idea is illustrated with a simple rigged-lottery example, and we outline how this idea may help to resolve a recent debate amongst cosmologists: is dark energy a cosmological constant, or has it evolved with time in some way? And how shall we decide, when the data are in?
\end{abstract}

\maketitle

%%%%%%%%%%%%%%%%%%%%%%%%%%%%%%%%%%%%%%%%%%%%
%% MAINMATTER
%%%%%%%%%%%%%%%%%%%%%%%%%%%%%%%%%%%%%%%%%%%%
\section{Introduction}
In Bayesian model selection, we have two or more competing hypotheses, $H_1$ and $H_2$, with each possibly containing different parameters $\theta_1$ and $\theta_2$. We wish to judge the plausibility of these two hypotheses in the light of some data $D$, and some prior information $I$, dropped hereafter for succinctness. Bayes' rule provides the means to update our plausibilities of these two models, to take into account the data $D$:
\begin{eqnarray}\label{bayes}
\frac{P(H_2|D)}{P(H_1|D)} = \frac{P(H_2)}{P(H_1)}\frac{P(D|H_2)}{P(D|H_1)}
= \frac{P(H_2)}{P(H_1)}\times\frac{\int p(\theta_1|H_1)p(D|\theta_1, H_1)\,d\theta_1}{\int p(\theta_2|H_2)p(D|\theta_2, H_2)\,d\theta_2}
\end{eqnarray}
Thus, the ratio of the posterior probabilities for the two models is the prior odds ratio times the evidence ratio.

If the various probabilities on the right-hand side of Equation~\ref{bayes} are a good description of our prior beliefs, then the posterior probabilities will encode justified conclusions based on the data. However, practical use of Equation~\ref{bayes} is often regarded with scepticism \citep{linder, liddle, efstathiou}. This is primarily because the probabilities on the right-hand side are difficult to specify without making ad hoc choices.

For reasons that are mostly historical, the prior distributions $p(\theta_1|H_1)$ and $p(\theta_2|H_2)$ for the parameters of each model are usually considered the most troubling. The prior model probabilities are often set to $1/2$, citing symmetry, and the sampling distributions are usually considered uncontroversial. However, in many real scientific applications, assigning priors is trivial compared to the job of assigning sampling distributions (Hogg, priv comm); i.e. modelling how the question of interest would affect our data.

While many Bayesians would assert that the dependence on subjective judgments exists because the result should actually depend on these judgments, it seems as though there ought to be ways to reduce the subjective influences in the prior probabilities and sampling distributions, even if they can never be entirely eliminated. In fact, this is the entire reason for using Bayes' rule in the first place \citep{ohagan}. Rather than simply looking at the data and then assigning a posterior distribution directly, we make use of one objective thing we actually know, Bayes' rule. In this paper, we discuss how the principle of maximum relative entropy (ME) \citep{caticha} can be used to further reduce, though not eliminate, the subjectivity of Bayesian inferences. The key requirement of this approach is that we must have a realistic probabilistic model of our prior beliefs about the data, i.e. our prior predictive distribution for the data must be modelled carefully.

\subsection{Publishing the Evidence}
\citet{skilling} recommends that whenever some data is analysed using a model $M_1$, the evidence $Z_1 = p(D|M_1) = \int p(\theta_1)p(D|\theta_1)d\theta_1$ be presented. This way, anyone proposing a different model $M_2$ can calculate their own evidence $Z_2$ and carry out model comparison with Equation~\ref{bayes} without the need to recalculate $Z_1$, which was published by the first author. This is good advice that has been taken by many in the astronomical community \citep{trotta, koopmans}, however, it is not the whole story. The plausibility of a model does not depend only on the evidence, it also depends on the prior probability (Equation~\ref{bayes}). A large evidence ratio can easily be cancelled by a tiny prior probability ratio and vice versa. The sure thing problem, discussed in Section~\ref{surething}, is simple and well-known example of this fact.

\section{A Sure Thing Problem}\label{surething}
Suppose a simple lottery is held, with tickets numbered from 1 to 1,000,000. Each ticket is sold to a different person. Consider a hypothesis $H_1$, which states that the lottery is fair, and thus the probability of any particular ticket winning is $10^{-6}$. The draw is carried out, producing the following data $D$: The winner of the lottery was ticket \#$263878$. Alice publishes a paper that reports this data, and proposes the fair lottery model $H_1$ to explain it. She presents the evidence $Z_1 = P(D|H_1) = 10^{-6}$.

Bob, a professional rival of Alice, reads her paper and proposes a different model, $H_2$: The lottery was not fair. It was rigged in order to make ticket \#$263878$ the winner. Bob writes a paper presenting the evidence $Z_2 = P(D|H_2) = 1$. Thus, he concludes, if $H_1$ and $H_2$ are initially equally plausible, the data makes $H_2$ a million times more plausible than $H_1$. Clearly, something is not quite right with this conclusion.

\subsection{Jaynes' Solution: Introduce extra hypotheses}
\citet{jaynes} resolves the sure thing paradox in the following way. When Bob does a model selection between $H_1$ and $H_2$ with $P(H_1) = P(H_2) = \frac{1}{2}$, he is implicitly stating that before getting the data, he would have predicted ticket \#$263878$ with a probability greater than 50 \%. Clearly, there is no way he could have known this before seeing the data. Actually, before observing the data, there were 999,999 other ``sure thing'' hypotheses that were on an equal footing with $H_2$. The correct analysis would involve a bigger hypothesis space containing 1,000,001 hypotheses: $H_1$, and the 1,000,000 sure thing hypotheses $\{S_1, S_2, ..., S_{1,000,000}\}$, where $S_{263878} \equiv H_2$. Bob should have assigned 1/2 of the prior probability to $H_1$ and divided the other 1/2 evenly amongst the $S$'s. Then, the prior probability of $H_2$ is $5 \times 10^{-7}$ and its posterior probability is $1/2$. This is the correct result; knowledge of the winning ticket number does not affect the plausibility of foul play. This argument resolves the sure thing problem by introducing a large number of alternatives into the hypothesis space, thus drastically reducing the prior probability of the particular sure thing hypothesis selected by the data. However, it is difficult to generalise this reasoning into more complicated scenarios where the principle of indifference cannot be used.

Before the lottery was drawn, Bob would have assigned a uniform predictive distribution for the data. His reanalysis ought to reflect this, if not by introducing extra sure thing models, then by downweighting $H_2$ somehow to reduce the spike it produces in the predictive distribution. 
While this is not the explicit motivation for entropic priors, it is a pleasant side effect, as we will show in the next section.

\section{Entropic Priors}\label{entropic}
In this section we introduce the notion of an entropic prior \citep{2004PhRvE..70d6127C, rodriguez}. Usually, Bayesian Inference is concerned with describing our knowledge in two stages: before taking into account the data, and then after taking into account the data. Bayes' rule is used to do this updating. Before taking into account the data, there is a prior distribution $p_1(\theta)$ and sampling distributions $p_1(D|\theta)$ for all $D$ and $\theta$. The reason for the subscript `1' will become clear later. By the product rule, this is equivalent to defining a {\it joint prior} on the product space of possible hypotheses and possible data:
\begin{equation}
p_1(\theta, D) = p_1(\theta)p_1(D|\theta)
\end{equation}
Here, the usual prior $p_1(\theta)$ (actually a marginal distribution) describes prior knowledge about $\theta$, and the sampling distributions $p_1(D|\theta)$ describe prior knowledge about how $\theta$ is related to the data $D$ that we plan to observe. The key point here is that before learning the data, we are uncertain both about the parameters {\it and} about the data: $p_1(\theta,D)$ should model this state of uncertainty.

In this paper, we will be concerned with describing uncertain knowledge about $(\theta, D)$, so we will be using probability distributions on the product space. We will start from a joint prior $p_0(\theta, D)$ and update this distribution {\it twice} to obtain the final joint posterior. We thus describe knowledge at three stages, defined below.

\begin{itemize}
\item Stage 0: Before we observe the data, or even know what sampling distributions are. However, the parameter space and the data space have been defined, as well as priors over these spaces. At stage 0, our knowledge is $p_0(\theta,D)$.
\item Stage 1: Also before we observe the data. However, we have now specified the sampling distributions $p(D|\theta)$ for all $\theta$ and $D$. At stage 1, our knowledge is $p_1(\theta,D)$.
\item Stage 2: We now have the data. Our knowledge is $p_2(\theta,D)$.
\end{itemize}

Updating from Stage 1 to Stage 2 is what we typically think of as Bayesian analysis. We prefer updating, rather than just writing down Stage 2 probabilities, because we get to use an objective updating rule, Bayes' rule. The idea behind entropic priors is to split up the process of assigning Stage 1 probabilities into two steps: Assigning Stage 0 probabilities, and then updating to Stage 1 using another objective updating rule, ME \citep{caticha}. There is a lot of confusion in the literature about the relationship between these two principles. However, there need not be any tension between them if it is understood that Bayes' rule is to be used when we learn about propositions built from those in the product space, such as `$D=42$' or `$\theta+D\leq 1$', whereas ME applies to propositions about {\it probability distributions on that space}\footnote{This raises a philosophical point, as most information is ultimately in the form of data. However, we may summarise the resultant effect of a large amount of external data as providing a constraint on our probability distributions.}, such as `$p(\theta, D)$ should be a Gaussian'.

\subsection{Updating from Stage 0 to Stage 1}
Say we have a Stage 0 joint prior, and we don't know the sampling distributions yet. Perhaps we haven't calibrated the instruments to see what kinds of output they typically produce. At this point our knowledge of $(\theta, D)$ will often be independent, such that taking data before learning about the experiment does not tell us anything {\it about the parameters} (it does tell us the data - and is therefore significant information in the product space. However, data is usually a nuisance parameter[!]). However, for generality we will allow dependence in the stage 0 distribution: $p_0(\theta,D) = p_0(\theta)p_0(D|\theta)$.

\begin{figure}
\includegraphics[scale=1.3]{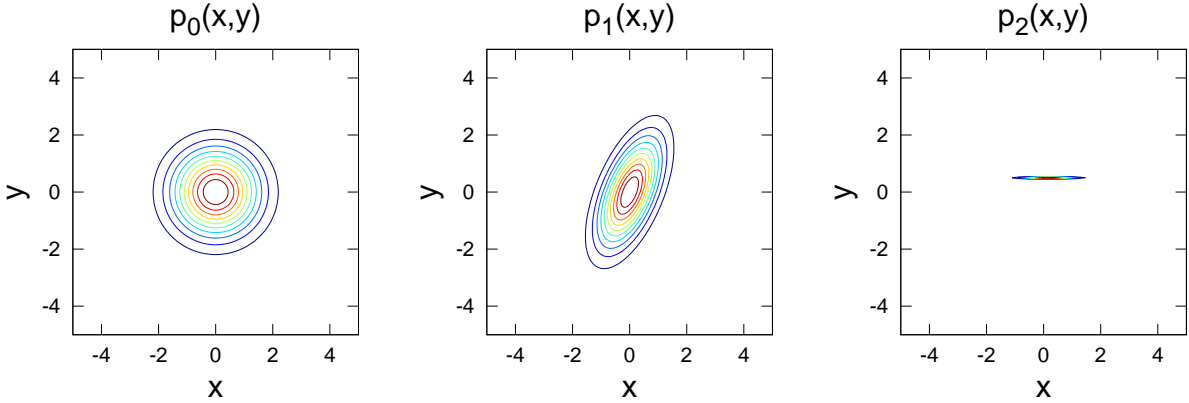}\caption{The basic idea behind entropic priors. An initially independent  $N(0,1)$ joint prior for two quantities $x$ and $y$ (to be thought of as ``parameters'' and ``data'' respectively) is updated once the sampling distributions $p(y|x) \,\forall y,x$ are known to be $p(y|x)\sim N(x,1)$. When the data are known (in this case, $y=0.5$), the joint distribution is updated again. This second updating is equivalent to the usual Bayesian process.\label{updating}}
\end{figure}

We then learn information in the form of a {\it constraint on allowable joint probability distributions}: the sampling distributions $p(D|\theta)$ for all $\theta$ and $D$ are given to us: $p(D|\theta) = f(D; \theta)$, where $f$ is a given function. We must adjust our joint distribution so that this constraint is satisfied. By the rules of probability any distribution of the form $p_1(\theta, D) = p_1(\theta)f(D; \theta)$ is allowed, and we have absolute freedom to vary $p_1(\theta)$ while still satisfying the constraint on the sampling distributions. However, there is a best choice for $p_1(\theta)$ \citep{caticha}: $p_1(\theta)$ should be chosen such that $p_1(\theta,D)$ is as close as possible to $p_0(\theta, D)$, i.e. we choose the $p_1(\theta)$ that maximises the relative entropy
\begin{eqnarray}
S = -\int \int p_1(\theta,D)\log\left(\frac{p_1(\theta,D)}{p_0(\theta,D)}\right)\,d\theta\, dD \\
= -\int \int p_1(\theta)p_1(D|\theta)\log\left(\frac{p_1(\theta)p_1(D|\theta)}{p_0(\theta)p_0(D|\theta)}\right)\,d\theta\, dD 
\end{eqnarray}
Differentiating with respect to each value of $p_1(\theta)$ (i.e. its value at each $\theta$) and setting to zero (with Lagrange multiplier term added):
\begin{eqnarray}
\frac{\partial}{\partial p_1(\theta)}\left(S - \lambda\left(\int p_1(\theta) \,d\theta - 1\right)\right) = 0
\end{eqnarray}
Carrying out this calculation gives:
\begin{eqnarray}\label{entprior}
p_1(\theta) \propto p_0(\theta)e^{S(D|\theta)}
\end{eqnarray}
where
\begin{eqnarray}
S(D|\theta) = -\int p_1(D|\theta)\log\left(\frac{p_1(D|\theta)}{p_0(D|\theta)}\right)\,dD
\end{eqnarray}
Thus, the marginal for $\theta$ after learning the sampling distributions, $p_1(D|\theta)$, is proportional to the original marginal $p_0(\theta)$, but multiplied by the exponential of the entropy of the corresponding sampling distribution relative to $p_0(D|\theta)$ (usually just $p_0(D)$). This process of updating from stage 0 to stage 1 using ME, and subsequently updating using data, is illustrated graphically in Figure~\ref{updating}.

Applying similar logic to model selection problems consists simply of applying the short-cut reasoning of the above paragraph: each hypothesis (model and its parameter value) gets its prior probability rescaled by a factor measuring the closeness of its predictions to our initial predictions. This makes it clear that invoking symmetry to assign $P(H_1)=P(H_2)=\frac{1}{2}$ is flawed: the symmetry may be broken as soon as we assign the sampling distributions.

\section{Sure Thing Problem: Entropic Priors Solution}
For the lottery problem, our knowledge about the data before getting it, and before the two models have been specified, is described by a uniform distribution over the integers from 1 to $10^6$: $p_0(D) = 10^{-6}$ for all $D$\footnote{This could be different if we tried to incorporate psychological theories as to which numbers would be more likely under the cheating scenario - perhaps `123456' would have its probability boosted -  but we will ignore this effect here.}. Because the two models haven't been specified yet, symmetry implies we must assign equal prior (marginal stage 0) probabilities $p_0(H_1) = p_0(H_2) = \frac{1}{2}$. The joint probabilities for the hypothesis and the data are therefore uniform and independent.

The next step is to incorporate more information and update our probabilities to stage 1. This information is not the data, but the specification of the sampling distributions: $p(D|H_1) = 10^{-6}$ for all $D$, and $p(D|H_2) = 1$ if $D=263878$ and zero otherwise. As explained in Section~\ref{entropic}, the priors for the two hypotheses should be reweighted according to the exponential of the entropy of their sampling distributions with respect to the original predictive distribution. These entropies are

\begin{eqnarray}
S(D|H_1) = -\sum_{i=1}^{10^6} 10^{-6} \log \frac{10^{-6}}{10^{-6}} = 0\\
S(D|H_2) = -1 \log \frac{1}{10^{-6}} = \log(10^{-6}) \\
\end{eqnarray}

Thus, the solution to the lottery problem is:
\begin{eqnarray}
\frac{p(H_2|D)}{p(H_1|D)} = \left(\frac{\frac{1}{2}}{\frac{1}{2}}\right) \times \left(\frac{10^{-6}}{1}\right) \times \left(\frac{1}{10^{-6}}\right) = 1
\end{eqnarray}
The three factors here are the Stage 0 odds ratio, the entropic correction factor, and the evidence ratio. The resulting conclusion is as it should be: knowing the winning lottery number provides no information about whether there is fraud or not. Usually, models that make sharp, correct predictions are favoured by Bayesian inference. In this example, this still occurs in the evidence ratio, but the entropic factor also {\it penalizes} $H_2$ by the same amount for being unjustifiably confident compared to our honest prior predictive distribution $p_0(D)$.

\section{Evolving Dark Energy}
The nature  of dark energy, thought  to be responsible  for causing the
observed  late-time accelerated expansion  of the  Universe \citep{1998AJ....116.1009R, 1999ApJ...517..565P}, is  a key driver of  many upcoming cosmological surveys and  instruments \citep{2009arXiv0901.0721A}. From a model selection point of view, one of the key questions is whether the equation of state  of dark energy, $w$, is exactly  equal to minus one for all  time or  whether it has  any temporal variation.   The former case  is equivalent  to Einstein's  cosmological constant,  or non-zero vacuum  energy, while any  variation from  this value,  however small, indicates very different  physics at play, such as  the existence of a primordial scalar  field, or other  even more exotic  possibilities \citep{2005PASA...22..315B}. Here,  model selection is really  the key goal; the  exact form of any evolution of  $w$ is less interesting than  simply being sure that it  does  evolve,  or  at  least  have  a  value  different  from  the cosmological constant. There has  been vigorous debate in  the literature on  how to best  answer this question \citep{linder, liddle, efstathiou}. Here we outline the contribution entropic priors can make to this debate; detailed analysis will be presented in a future contribution.

We will consider four different models for how the dark energy equation of state has varied throughout the universe's history. Typically this is described as $w$ varying as a function of $a$, the scale factor.
\begin{center}
\begin{tabular}{ll}
{$H_0$: \it Cosmological Constant $\Lambda$}: & $w(a) = -1$\\
{$H_1$: \it Constant, but non-$\Lambda$}: & $w(a) = w_0$\\
{$H_2$: \it Simple Evolving}: & $w(a) = w_0 + (1-a)w_a$\\
{$H_3$: \it Complex Evolving}: & $w(a) = $ \textnormal{Some model with many parameters}
\end{tabular}
\end{center}
Observations of type Ia supernovae, particularly how their apparent brightness decreases with redshift, is a strong probe of $w$ \citep{2008ApJ...686..749K, union}. The idea is to test the four models, given such data \citep{efstathiou}, and to forecast the informativeness of proposed future missions \citep{2009arXiv0901.0721A}. However, not all of the models are physically well-motivated: e.g. $H_0$ arises naturally from General Relativity, $H_1$ and $H_2$ are ad hoc ``simple models'', and $H_3$ expresses gross ignorance. Therefore, while it is fair to assign a large probability to $H_0$ at Stage 1, it does not {\it automatically} make sense to share the remaining probability evenly amongst $H_1$-$H_3$. The reason for this is that $H_1$, $H_2$ and $H_3$ may imply quite different predictive distributions for the data. If we build a $p_0(D)$ that we trust, then the simple models $H_1$ and $H_2$ may be downgraded in prior probability solely because they make predictions that are too confident\footnote{The chance that an unknown function just happens to be a straight line is usually quite small, unless you have very good prior reasons to expect a straight line.}. Of course, if, in the course of building $p_0(D)$, we explicitly think about the predictions of the $H$'s, then this will not occur - entropic priors do not magically generate information. What they do is implore us to think about $p(D)$ when assigning priors, a key sanity check that is often overlooked.

\section{Conclusions}
Bayesian model selection is a difficult task, both computationally and philosophically. If we are not careful, we can obtain misleading results. The idea presented in this paper, of assigning a realistic predictive distribution for the data and then penalizing models whose predictions differ from it, should assist in making Bayesian model selection analyses more reliable.

%%
%% End of file `template-6s.tex'.
%%%%%%%%%%%%%%%%%%%%%%%%%%%%%%%%%%%%%%%%%%%%%%%%
%% BACKMATTER
%%%%%%%%%%%%%%%%%%%%%%%%%%%%%%%%%%%%%%%%%%%%%%%%

\begin{theacknowledgments}
BJB would like to thank the following for valuable discussion: David W. Hogg, John Skilling, Ariel Caticha, Adom Giffin, Iain Murray, Phil Marshall and Geraint Lewis. I would also like to thank Georgina Wilcox for introducing me to \texttt{mercurial}, and Tommaso Treu for supporting my trip to Oxford, Mississippi.
\end{theacknowledgments}

\end{document}